\documentclass[twocolumn,prd,nofootinbib,showpacs,preprintnumbers]{revtex4}
\usepackage{graphicx}
\usepackage{color}
\usepackage{amsmath,amssymb,amsfonts,dsfont}
\usepackage{multirow}
\usepackage{sistyle}
\usepackage{ulem}

\pdfoutput=1 

\newcommand{\lsim}{\mathrel{\mathop{\kern 0pt \rlap
  {\raise.2ex\hbox{$<$}}}
  \lower.9ex\hbox{\kern-.190em $\sim$}}}
\newcommand{\gsim}{\mathrel{\mathop{\kern 0pt \rlap
  {\raise.2ex\hbox{$>$}}}
  \lower.9ex\hbox{\kern-.190em $\sim$}}}

\def \aap  {A\&A}

\def \apjs  {ApJS}
\def \apjl  {ApJL}

\begin{document}

\preprint{TTK-20-03}

\title{Contribution of pulsars to cosmic-ray positrons in light of recent observation of inverse-Compton halos}

\author{Silvia Manconi}\email{manconi@physik.rwth-aachen.de}
\affiliation{Institute for Theoretical Particle Physics and Cosmology, RWTH Aachen University, Sommerfeldstr.\ 16, 52056 Aachen, Germany}
\author{Mattia Di Mauro,}\email{mdimauro@slac.stanford.edu}
\affiliation{NASA Goddard Space Flight Center, Greenbelt, MD 20771, USA}
\affiliation{Catholic University of America, Department of Physics, Washington DC 20064, USA}
\author{Fiorenza Donato}\email{donato@to.infn.it}
\affiliation{Dipartimento di Fisica, Universit\`a di Torino, via P. Giuria 1, 10125 Torino, Italy}
\affiliation{Istituto Nazionale di Fisica Nucleare, Sezione di Torino, Via P. Giuria 1, 10125 Torino, Italy}

\begin{abstract}
The hypothesis that pulsar wind nebulae (PWNe) can significantly contribute to the excess of the positron ($e^+$) cosmic-ray flux 
has been consolidated after the observation of a $\gamma$-ray emission at TeV energies of a few degree size around Geminga and Monogem PWNe, 
and at GeV energies for Geminga at a much larger extension.
The $\gamma$-ray halos around these PWNe are interpreted as due to electrons ($e^-$) and $e^+$ accelerated and escaped by their PWNe, and inverse Compton scattering  low-energy photons of the interstellar radiation fields.  
The extension of these halos suggests that the diffusion around these PWNe is suppressed by two orders of magnitude with respect to the average in the Galaxy. 
We implement a two-zone diffusion model for the propagation of $e^+$ accelerated by the Galactic population of PWNe.
We consider pulsars from the ATNF catalog and build up simulations of the PWN Galactic population. 
In both  scenarios, we find that within a two-zone diffusion model, the total contribution from PWNe and secondary $e^+$ is at the level of AMS-02 data, for an efficiency of conversion of the pulsar spin down energy in $e^\pm$  of $\eta\sim0.1$.  For the simulated PWNe, a $1\sigma$ uncertainty band is determined, which is of at least one order of magnitude from 10 GeV up to few TeV. 
The hint for a decreasing $e^+$ flux at TeV energies is found, even if it is strongly connected to the chosen value of the radius of the low diffusion bubble around each source. 
\end{abstract}

\maketitle

\section{Introduction}
\label{sec:intro}
Evidence for  an excess of the positron ($e^+$) component of cosmic rays (CRs)  has been first measured by Pamela \cite{Adriani:2013uda} and {\it Fermi}-LAT \cite{2012PhRvL.108a1103A}, 
and then confirmed  with unprecedented precision by AMS-02 \cite{PhysRevLett.122.041102}.
The excess refers to the observed flux of $e^+$ above 10 GeV, which cannot be explained  by spallation reactions of CRs with the Interstellar Medium (ISM) alone  \cite{Delahaye:2008ua,Diesing:2020jtm}. 
Several explanations have been proposed in the literature, invoking $e^+$ accelerated from pulsar wind nebulae (PWNe) (see, {\it e.g.}, \cite{Hooper:2008kg,Mlyshev:2009twa,Gaggero:2013nfa,2014JCAP...04..006D,Boudaud:2014dta,Cholis:2018izy}), supernova remnants (SNR) (see, {{\it e.g.}, \cite{2009PhRvL.103e1104B,Mertsch:2014poa}), produced by dark matter particle interactions (see, {\it e.g.}, \cite{2010JCAP...01..009I,2016JCAP...05..031D}), 
or modifications of the standard picture of CR propagation in the Galaxy \cite{Tomassetti:2015cva,Lipari:2018usj}. 
The intense radiative losses suffered by high energetic $e^\pm$ require a hypothetical primary source of CR $e^+$ to be local, {\it i.e.}~at few kpc from the Earth. 
For this reason, the scenario in which a nearby source dominates the observed flux has received particular interest, {\it e.g.}~for the two most powerful PWNe near the Earth, Geminga and Monogem \cite{2009APh....32..140G,Linden:2013mqa,Manconi:2016byt}. 

The recent observation of a $\gamma$-ray emission at TeV energies of a few degree  size  reported by HAWC \cite{Abeysekara:2017science} and Milagro \cite{2009ApJ...700L.127A} in the direction of Geminga and Monogem PWNe further supports the idea that these objects might be the sources of primary $e^+$ in our Galaxy.  
In fact, the $\gamma$-ray halos detected around Geminga and Monogem are interpreted as due to electrons ($e^-$) and $e^+$ accelerated, and escaped, by their PWNe and inverse Compton scattering (ICS) low-energy photons of the interstellar radiation fields.
The presence of a $\gamma$-ray halo around Geminga has been recently confirmed also at GeV energies  with an analysis of {\it Fermi}-LAT data above 8~GeV \cite{DiMauro:2019yvh}. 
At these energies, the size of extension is much larger, and reaches about 15 degrees at 10 GeV.
In general, combined GeV and TeV observations of such halos further constrain the properties of the accelerated $e^\pm$, such as the spectral index of the emission \cite{DiMauro:2019yvh,Shao-Qiang:2018zla}. 
Moreover, the extension of the Geminga and Monogem halos suggests that the diffusion around these PWNe is suppressed by two orders of magnitude with respect to the value fitted to the AMS-02 CR nuclei data  (see, {\it e.g.}, \cite{Kappl:2015bqa,Genolini:2015cta,Genolini:2019ewc}). 
The inferred diffusion coefficient is  about $10^{26}$~cm$^2$/s at 1 GeV \cite{Abeysekara:2017science,DiMauro:2019yvh}.

The observation of the ICS halos around Geminga and Monogem at different energies has provided key information about the acceleration mechanisms of $e^{\pm}$ from PWNe and their propagation in the Galactic environment.
Following the HAWC, Milagro and {\it Fermi}-LAT observations, several authors have studied the flux of $e^+$ from PWNe, and have drawn conclusion on the contribution of this source population to the $e^+$ excess \cite{Hooper:2017gtd,Abeysekara:2017science,Shao-Qiang:2018zla,Tang:2018wyr,Fang:2018qco,DiMauro:2019yvh}. 
When using the low diffusion found around Geminga and Monogem PWNe for propagating particles in the entire Galaxy, the contribution from these two pulsars  
to the $e^+$ flux is found to be negligible \cite{Abeysekara:2017science}, and other sources are needed in order to explain the CR $e^\pm$ data \cite{Lopez-Coto:2018ksn}. 
In order to account for the inefficient zone of propagation found around Geminga and Monogem, a more detailed phenomenological two-zone diffusion model has been introduced in Ref. \cite{Tang:2018wyr,Profumo:2018fmz}.
As detailed in Ref.~\cite{DiMauro:2019yvh} (see also \cite{Shao-Qiang:2018zla}), the analysis of the flux and morphology of the Geminga ICS halo from GeV to TeV energies suggests that, in a two-zone diffusion model, this source contributes at most $10\%$ to the $e^+$ excess.
The origin of these inefficient diffusion bubbles around Galactic PWNe have been also studied, although a comprehensive description is still under debate, in particular for very old objects such as Geminga \cite{2018arXiv180709263E,Kun:2019sks,Liu:2019zyj}. 

Further evidences that the ICS halos might be a general feature of all Galactic PWNe have been recently discussed \cite{Sudoh:2019lav,DiMauro:2019hwn}.  
In particular, in our Ref.~\cite{DiMauro:2019hwn}, we presented a systematic study of PWNe detected by HESS. 
Ranking them according to the ICS halo flux a 10~TeV, we found that for the  brightest sources, indeed a model for the ICS halo describes better the observed $\gamma$-ray emission with respect to a simple geometrical 2-d Gaussian model. 
We provided, for about 20 sources spanning ages from 5~kyr to 240~kyr, a measurement of the diffusion coefficient at TeV energies around these objects. Similarly to the case of Geminga, we found that this is systematically smaller by about two orders of magnitude than the value considered to be the average in the Galaxy. 

In this paper we present an extensive study aimed at quantifying the total contribution of Galactic PWNe to the $e^+$ flux observed at Earth.
We will include in our calculations an inefficient diffusion zone present around each Galactic PWNe, as suggested by our analysis of candidate ICS $\gamma$-ray halos \cite{DiMauro:2019hwn}. 
To quantify the effect of these observations, we assume a two-zone diffusion model for the propagation of $e^+$ from each source \cite{Tang:2018wyr}, and we vary the value of the diffusion coefficient and the radius of the inefficient bubble around the values most likely to be representative for all the Galactic PWNe. 
We will present results both for the observed Galactic pulsars present in the ATNF catalog \cite{2005AJ....129.1993M}, and for simulations of PWNe with a spatial distribution following the Galactic spiral arms. 

The paper is organized as follows. 
In Sec.~\ref{sec:model}  we outline the model from the production and propagation of $e^{\pm}$ from PWN to the Earth. We also describe how we implement the ATNF catalog parameters as well as how we generate the simulations of Galactic PWNe. 
In Sec.~\ref{sec:results} we discuss our results for the $e^+$ arriving at the Earth both for the cataloged sources as well as for our  PWN simulations. 
Finally, in Sec.~\ref{sec: conclusions} we draw our conclusions.

\section{Modeling the $e^{\pm}$ flux at Earth from PWNe}
\label{sec:model} 

We recall here the basics for the model of $e^{\pm}$ flux from PWNe, based on the formalism detailed in Ref. \cite{DiMauro:2019yvh}. 
In the first part of this section we discuss the injection spectrum of $e^{\pm}$ from PWNe, and the propagation of accelerated particles in the Galaxy, under the one-zone or two-zone propagation models. 
Then we explain in Sec.~\ref{sec:atnf} and Sec.~\ref{sec:simulations} our assumptions for the spatial distribution and properties of the Galactic pulsar population.

PWNe are thought to accelerate and inject $e^{\pm}$ in the ISM up to very high energies (see, {\it e.g.}, \cite{1996ApJ...459L..83C,Amato:2013fua,Gaensler:2006ua}).
The rotation of the pulsar induces an electric field that extracts $e^-$ from the star surface. 
These $e^-$ lose energy via curvature radiation while propagating far from the pulsar along the magnetic field lines, and the very-high-energy emitted photons create a wind of $e^{\pm}$ pairs in the intense neutron star magnetic field. 
During the free expansion, the pulsar wind meets the SNR  ejecta expanding in the ISM, creating a forward and reverse shock.
The latter constitutes a termination shock, and its bulk energy is dissipated into a relativistically, magnetized fluid, which shines as a PWN.
The $e^{\pm}$ pairs produced in the pulsar magnetosphere reach the termination shock, and a relatively large fraction (up to few tens of percent) of the wind bulk energy can be converted into accelerated $e^{\pm}$ pairs. 
They then radiate into a photon spectrum extending from radio frequencies to TeV $\gamma$-rays, through synchrotron and ICS processes \cite{Gaensler:2006ua,Bykov:2017xpo}.

We consider a model in which $e^{\pm}$ are continuously injected at a rate that follows the pulsar spin-down energy $W_0$. This  scenario is indeed required to generate the TeV  photons  detected by Milagro and HAWC for Geminga and Monogem \cite{Yuksel:2008rf,Abeysekara:2017science,DiMauro:2019yvh}. 
A common alternative is the burst-like scenario, according to which all the particles are emitted from the source at a time equal to the age of source $T$. 
In the continuous injection model, the injection spectrum $Q(E,t)$ at a time $t$ can be described as:
 \begin{equation}
 Q(E, t)= L(t) \left( \frac{E}{E_0}\right)^{- \gamma_e} \exp \left(-\frac{E}{E_c} \right) ,
 \label{eq:Q_E_cont}
\end{equation}
where the magnetic dipole braking $L(t)$ (assuming a magnetic braking index of 3) is defined as:
 \begin{equation}
 L(t) = \frac{L_0}{\left( 1+ \frac{t}{\tau_0} \right)^{2} }.
 \label{eq:Lt}
\end{equation}
The cutoff energy $E_c$ is fixed to $10^3$~TeV and the  characteristic pulsar spin-down timescale  $\tau_0=12$~kyr, following \cite{Abeysekara:2017science,DiMauro:2019yvh}.   
The normalization of the power law is fixed to $E_0 =1$~GeV.
The spectral index $\gamma_e$ of accelerated particles is uncertain, and may vary significantly among different PWNe \cite{Mlyshev:2009twa,Gaensler:2006ua}.  In what follows we consider different possibilities, varying $\gamma_e$ in the range $[1.4, 2.2]$.
The total energy emitted by the source in $e^{\pm}$ is given by:
 \begin{equation}
 E_{{\rm tot}} = \eta W_0 =\int_0^{T} dt \int_{E_{1}}^{\infty} dE E Q(E,t),
 \label{eq:Q_E_cont}
\end{equation}
where we fix $E_1 = 0.1$ GeV \cite{Buesching:2008hr,Sushch:2013tna}. 
The parameter $\eta$ encodes the efficiency of conversion of the spin-down energy into $e^\pm$ pairs.
$W_0$ can be computed from cataloged quantities as the pulsar age $T$, 
the decay time $\tau_0$, and the spin-down luminosity $\dot{E}$:
\begin{equation}
 W_0 = \tau_0 \dot{E} \left( 1+ \frac{T}{\tau_0} \right)^2\,.
 \label{eq:W0PWN}
\end{equation}
The \textit{actual} age $T$ and the  \textit{observed} age $t_{\rm{obs}}$ are related by the source distance $d$ by $T = t_{\rm obs} + d/c$.

In the continuous injection scenario and with a homogeneous diffusion in the Galaxy, the $e^\pm$ number density per unit volume and energy 
 $\mathcal{N}_e(E,\mathbf{r},t)$ of $e^\pm$ at an observed energy $E$, a position $\mathbf{r}$ in the Galaxy, and  time $t$ is given by  \cite{Yuksel:2008rf}:
 \begin{eqnarray}
\label{eq:N_cont}
  \mathcal{N}_e(E,\mathbf{r},t) &=& \int_0^{t} dt'\, \frac{b(E_s)}{b(E)} \frac{1}{(\pi \lambda^2(t',t,E))^{\frac{3}{2}}} \times \nonumber \\
  && \times  \exp\left({-\frac{|\mathbf{r} -\mathbf{r_{s}} |^2}{ \lambda(t',t,E)^2}}\right)Q(E_s,t'),
\end{eqnarray}
where the integration over $t'$ accounts for  the PWN releasing $e^\pm$ continuously in time.
The  energy $E_s$ is the initial energy of $e^\pm$ that cool down to $E$ in a {\rm loss time} $\Delta \tau$:
\begin{equation}
 \Delta \tau (E, E_s) \equiv \int_{E} ^{E_s} \frac{dE'}{b(E')} = t-t_{{\rm obs}} .
\end{equation}
The $b(E)$ term is the energy loss function, $\mathbf{r_{s}}$ indicates the source position, and $\lambda$ is the typical propagation scale length defined as:
\begin{equation}
\label{eq:lambda}
 \lambda^2= \lambda^2 (E, E_s) \equiv 4\int _{E} ^{E_s} dE' \frac{D(E')}{b(E')},
\end{equation} 
with $D(E)$ the diffusion coefficient.
The $e^\pm$ energy losses include ICS off the interstellar radiation field, 
and the synchrotron emission on the Galactic magnetic field. The  interstellar photon populations
at different wavelengths have been taken from \cite{Vernetto:2016alq}. The Galactic magnetic field intensity 
has been assumed $B=3.6\; \mu$G, as resulting from the sum (in quadrature) of the regular and turbulent components \citep{2007A&A...463..993S}. 
For further details on our treatment of the propagation in the Galaxy we address to \cite{Manconi:2016byt,Manconi:2018azw} (and refs. therein). 
\\
The flux of $e^\pm$ at Earth from a source is given by:
\begin{equation}
 \Phi_{e^\pm}(E) = \frac{c}{4\pi} \mathcal{N}_e(E,r=d,t=T).
 \label{eq:flux}
\end{equation}
We will assume, as a benchmark case, the propagation in the Galaxy as derived in Ref. \cite{Kappl:2015bqa} (hereafter K15) (see also \cite{Manconi:2016byt}). 

Recent results  \cite{Abeysekara:2017science,DiMauro:2019yvh} suggest that the diffusion coefficient around Geminga and Monogem PWNe is $\sim 10^{26}$~cm$^2$/s at 1 GeV, {\it i.e.}~about two orders of magnitude smaller than the value derived for the entire Galaxy through a fit to AMS-02 CR nuclei data \cite{Kappl:2015bqa,Genolini:2015cta,Genolini:2019ewc}.
A  phenomenological description  for this discrepancy proposes  a two-zone diffusion model, where the region of low diffusion is contained around the source, and delimited by an empirical radius $r_b$ \cite{Profumo:2018fmz,Tang:2018wyr}. 
 We stress here that our main purpose is to derive the consequences of the presence of such inefficient diffusion zones around Galactic PWNe using such phenomenological description, while no attempt is made to provide a detailed theoretical interpretation of this phenomenon.
The inhibition of diffusion near pulsars has been recently discussed {\it e.g.} in Ref. \cite{2018arXiv180709263E}, where a possible theoretical interpretation is provided.
In this paper we implement the two-zone diffusion model as in Ref. \cite{Tang:2018wyr,DiMauro:2019yvh}, for which the diffusion coefficient is defined as:
\begin{eqnarray}
\label{eq:conddm}
D(E,\rho) =  
\left\{
\begin{array}{rl}
& D_0 (E/1{\rm \,GeV})^\delta {\rm \;for\;} 0 < \rho < r_b, \\
& D_2 (E/1{\rm \,GeV})^\delta {\rm \;for\;} \rho \geq r_b.
\end{array}
\right.
\label{eq:Diff}
\end{eqnarray}
where $\rho$ is here the distance from the center of the pulsar.

In the two-zone diffusion model, the solution to the diffusion equation is modified with respect to Eq.~\ref{eq:N_cont}, and  the $e^{\pm}$ density takes the form \cite{Tang:2018wyr}:
 \begin{equation}
\label{eq:N_cont_2z}
  \mathcal{N}_e(E,\mathbf{r},t) = \int_0^{t} dt_0 \frac{b(E(t_0))}{b(E)} Q(E(t_0)) \mathcal{H}(\mathbf{r},E)\,. 
\end{equation}
The term $\mathcal{H}(\mathbf{r},E)$ is defined as:
\begin{eqnarray}
&& \mathcal{H}(\mathbf{r},E) =  \frac{\xi(\xi+1)}{(\pi \lambda_0^2)^{\frac{3}{2}} [2 \xi^2 {\rm erf}(\epsilon) - \xi(\xi-1) {\rm erf}(2\epsilon) + 2 {\rm erfc}(\epsilon)]} \nonumber \\
&&  \left\{
\begin{array}{rl}
 e^{({-\frac{\Delta r^2}{ \lambda_0^2}})} + \left(\frac{\xi-1}{\xi+1}\right) \left(\frac{2 r_b}{r}-1\right) e^{({-\frac{(\Delta r - 2 r_b)^2}{ \lambda_0^2}})},  0 < r < r_b \\
 \left( \frac{2\xi}{\xi+1} \right) \left[ \frac{r_b}{r} +\xi\left( 1 - \frac{r_b}{r} \right) \right] e^{( -[{\frac{(\Delta r - r_b)}{ \lambda_2}} + \frac{r_b}{\lambda_0} ]^2 )}, r \geq r_b,
\end{array}
\right.
\label{eq:Diff}
\end{eqnarray}
where $\Delta r = |\bf{r}-\bf{r_s}|$, $\xi=\sqrt{D_0/D_2}$, $\lambda_0$ and $\lambda_2$ are the typical propagation lengths for $D_0$ and $D_2$ (see Eq.~\ref{eq:lambda}), and $\epsilon = r_b/\lambda_0$.
We note that for $D_0=D_2$,   or assuming $r_b \gg \rho$,  the solution for two-zone diffusion model in Eqs.~\ref{eq:N_cont_2z} becomes Eq.~\ref{eq:N_cont}, which is valid indeed for a one-zone model.

According to  the results of \cite{Abeysekara:2017science,DiMauro:2019yvh,DiMauro:2019hwn}  the radius $r_b$ of the low-diffusion zones is at least $r_b>30$~pc. 
Assuming that around each Galactic pulsar there is a bubble of radius $r_{\rm b}$ in which $D_0$ is smaller with respect to the mean value in the Galaxy, the fraction of the Milky Way propagation volume occupied by those regions may be written as \cite{Hooper:2017tkg}:
\begin{equation}
 f\sim \frac{N_{\rm ICS} \times 4/3 \times \pi r_{\rm ICS}^3}{\pi R_{MW}^2 \times 2 z_{\rm MW}}=
\end{equation}
\begin{equation*}
 \sim 0.007 \left(\frac{r_b}{30 \rm{pc}} \right)^3 \left( \frac{\dot{N}_{\rm PSR}}{0.03\rm{yr}^{-1}}\right) \left( \frac{\tau_{ICS}}{10^6 \rm{yr}} \right) 
 \left( \frac{20 \rm{kpc}}{R_{\rm MW}} \right) \left( \frac{200 \rm{pc}}{z_{\rm MW}}\right)
\end{equation*}
where $N_{\rm ICS}$ is the number of ICS halos  at a given time in the Galaxy, and $R_{\rm MW}$ and $z_{\rm MW}$ are the radius and half-width of the Milky Way disk, respectively.   Taking $\dot{N}_{\rm PSR}$ as the pulsar birth rate, and $\tau_{ICS}$ as the time for such region to persist, we can write $N_{\rm ICS}=\dot{N}_{\rm SN}\times \tau_{ICS}$. 
The fraction $f$ is very sensitive to  $r_{\rm b}$. 
Assuming $\dot{N}_{\rm PSR}$ to be $1.4$ per century \cite{2004IAUS..218..105L},  we obtain $f\sim 0.0029$ for $r_{\rm b}=30~$pc. 
If $r_{\rm b}=30~$pc, this fraction is thus negligible.
For $r_{\rm b}=60~$pc, we obtain$f\sim0.023$, and for $r_{\rm b}=90~$pc and $r_{\rm b}=120~$pc, $f\sim0.078$ and  $f\sim0.18$, respectively. 
For $r_{b}>120~$pc the fraction of the Galactic propagation volume occupied by those regions is not negligible anymore. 
This can raise up to $40\%$, by increasing the pulsar birth rate to $\dot{N}_{\rm PSR}=0.03$. 
A different approach in the propagation of CRs in the Galaxy may thus be needed for very large values of $r_b$, in order to account for the global effect of the low-diffusion zones originated from all the PWNe on the propagated cosmic-rays. 
We leave this study to future work.  
In what follows we explore values of $r_b$ in the range [30, 120]~pc. 
The $e^+$ flux from each PWNe is computed by assuming a two-zone diffusion model, where a region $\rho<r_b$ of low-diffusion is considered around each PWNe, see Eqs.~\ref{eq:Diff},\ref{eq:N_cont_2z}. 

As for the number, spatial distribution and energetics of the Galactic pulsar population we follow two approaches, which we detail below. 
First (Sec.~\ref{sec:atnf}), we consider the observed Galactic PWNe. We use for this scope the list of pulsars reported in the ATNF catalog \cite{2005AJ....129.1993M}, similarly to what is done in Ref. \cite{Manconi:2016byt,Manconi:2018azw}.
In the second approach (Sec.~\ref{sec:simulations}) we instead consider mock catalogs of PWNe, produced by running simulations with a spatial distribution following the Galactic spiral arms, and pulsar properties ({\it e.g.}, age and spin-down luminosity) shaped on the observed Galactic pulsars. 

\subsection{ATNF pulsars}
\label{sec:atnf}
We use the ATNF catalog v1.57 \cite{2005AJ....129.1993M}, where 2627 sources are listed.
This is the most complete catalog of pulsars detected from radio to $\gamma$-ray energies, and is continuously updated to new discoveries.
To compute the $e^+$ flux, we implement the cataloged distance $d$, age $T$, and spin-down luminosity $\dot{E}$ given for each source. 
We consider only sources with an available value for these parameters. 
We select sources with ages between $50$~kyr and $10^5$~kyr, which decreases the sample to 1588 sources. 
The lower limit at $T<50$~kyr excludes sources for which the accelerated $e^\pm$ might be  still confined in the PWNe. In order  to compute the $e^+$ flux, we also need  a value for $\gamma_e$ (Eq.~\ref{eq:Q_E_cont}) and for the efficiency $\eta$ (Eq.~\ref{eq:Q_E_cont}). We will explore different scenarios, in which all the ATNF pulsars  share a common spectral index and efficiency, or where these values are drawn from a uniform distribution, see Sec.~\ref{sec:res_cat}. 
The spatial distribution of the pulsars in the ATNF is highly concentrated among few kpc, and thus this sample  is highly incomplete on a Galactic scale. 
Nevertheless, the typical propagation scale of high-energetic $e^\pm$ is limited to few kpc, as they suffer severe radiative losses. 
We thus expect that the sources listed in the ATNF catalog should contribute to the large majority of the $e^+$ flux observed at Earth.
Galactic distributions of pulsars which correct for this incompleteness have been computed in {\it e.g.} Ref.~\cite{2004IAUS..218..105L}, and are used as outlined in the next subsection.

\subsection{Simulation of Galactic pulsars }
\label{sec:simulations}
We generate simulations of Galactic PWNe using the  source population models implemented in the Python module \texttt{gammapy.astro.population} \cite{2015ICRC...34..789D}.
Using this module, we produce mock catalogs of Galactic pulsars, based on different assumptions for their spatial distribution, and with observed energetics. 
In what follows we list the main properties of these simulations, while we address to the code documentation\footnote{\url{https://docs.gammapy.org/dev/astro/population/index.html}} for any further detail. 

The mock catalogs provide the values of $T, d, \dot{E}$ and $\tau_0$ for each simulated source as follows. 
The total number of sources in each simulation is defined as $N_{\rm PSR} = t_{\rm max}\times \dot{N}_{\rm PSR}$, where $t_{\rm max}$ is the maximum simulated age and $\dot{N}_{\rm PSR}$ is the pulsar birth rate. 
Different estimates for the Milky Way pulsar birth rate $\dot{N}_{\rm PSR}$ range from one to four per century \cite{Keane:2008jj,2004IAUS..218..105L,2006ApJ...643..332F}. 
We here assume the maximum age of the sources to be $t_{\rm max}=10^{7}$~yr, and $\dot{N}_{\rm PSR}=0.01$~yr$^{-1}$. Accordingly, the simulation assigns to each mock pulsar a certain $T$. 
The value of $\dot{N}_{\rm PSR}$ acts as a global normalization for the cosmic-ray $e^+$ flux, and is degenerate with  $\eta$.

For each simulation, the radial distribution of sources is taken from the Lorimer profile  \cite{2004IAUS..218..105L}. 
In addition, we account for the spiral arm structure of our Milky Way according to the model of Ref.~\cite{2006ApJ...643..332F} (see their Table 2 for the spiral arm parameters).  
The further properties of the mock pulsars are drawn according to Ref. \cite{2006ApJ...643..332F}, see {\it e.g.} their Sec. 3.8. 
In these models, once the pulsar period $P_{\rm mean}$ and magnetic field $\log(B_{\rm mean})$ are defined, the parameters which are useful for the computation of the source-term for the cosmic-ray $e^+$ production, such as the distribution of their spin-down luminosities at birth or of $\tau_0$, are computed by modeling their time evolution.  
 As shown in Ref.~\cite{2006ApJ...643..332F}, the properties of observed sources in the ATNF catalog are reproduced by an initial normal distribution of pulsar periods - with $P_{\rm mean}=0.3$~s and with standard deviation of $P_{\rm std}=0.15$~s -  and  magnetic field 
 - with $\log(B_{\rm mean})=12.05$~G and  $\log(B_{\rm std})=0.55$~G. 
  The values for the spin-down energy at birth are then evolved for each source, to obtain the present spin down energy as $\dot{E}=\dot{E_0} (1+T/\tau_0)^{-2}$ \cite{Gaensler:2006ua,2011ASSP...21..624B}.
We simulate the case for which $\gamma_e$ and  $\eta$ have the same value for each mock pulsar, or they are drawn from a uniform distribution.

\begin{figure}[t]
\includegraphics[width=0.6\textwidth]{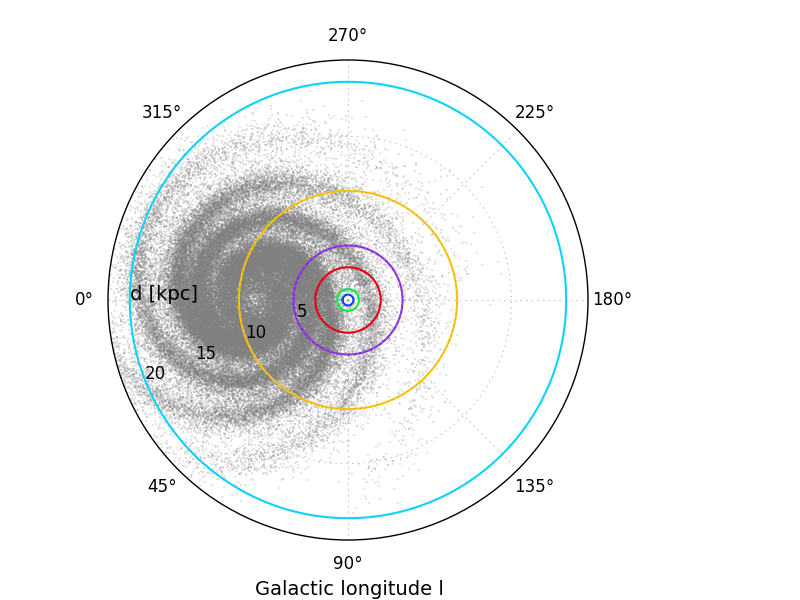}
\caption{Spatial distribution of simulated pulsars in one illustrative realization. The Earth is at the center of the plot, while the Galactic center is at $d=8.5$~kpc, $l=0$. 
The grey points indicates the position of each simulated pulsar. The concentric rings illustrate the distance rings we use to separate the contribution of simulated sources at different distances, see text for details.}
\label{fig:ringssim}
\end{figure}
In Fig.~\ref{fig:ringssim} we show the spatial distribution of pulsars in one illustrative simulation.
Concentric rings are drawn for iso-distances from the Earth of $0.5$~kpc (blue), $1$~kpc (green), $3$~kpc (red), $5$~kpc (purple), $10$~kpc (yellow) and $20$~kpc (cyan). 
The figure shows the characteristics distribution of sources along spiral arms. At distances close to the Earth, where sources contribute more to the $e^+$ flux, the pulsar density is smaller than at other distances, where spiral arms are located.

\section{Results for the $e^{\pm}$ flux at the Earth}
\label{sec:results}
In this paper we want to quantify the amount of $e^+$ arriving at Earth from all the Galactic pulsars, assuming that each source
is surrounded by a low diffusion bubble. We first evaluate the contribution from all the cataloged sources. Then, considering the 
possible incompleteness of the ATNF catalog, we extend our analysis to simulated Galactic pulsar populations. 

\subsection{Results for ATNF cataloged pulsars}
\label{sec:res_cat}
\begin{figure}[t]
\centering
\includegraphics[width=0.5\textwidth]{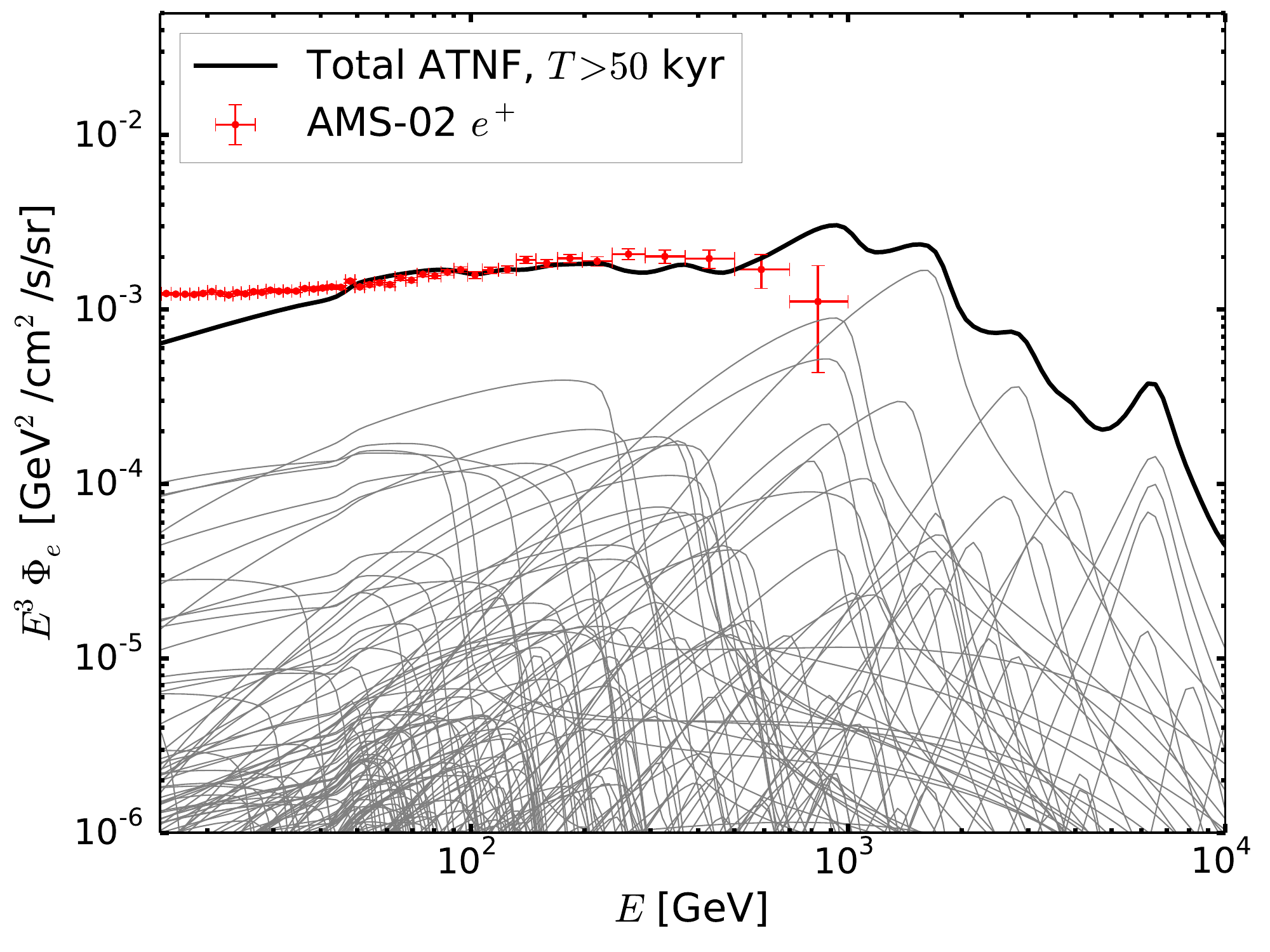}
\caption{ Predictions for the $e^+$ flux at the Earth from all the ATNF pulsars with $T>50$~kyr. 
The grey lines represent the contribution of each source, while the black line is their sum. 
The phenomenological two-zone diffusion model around the pulsar is implemented with $r_b=120$~pc and $D_0=7.8\times 10^{25}$ cm$^2$/s, 
while for the Galactic propagation model we have used  $D_{\rm K15}$. The $e^\pm$ are continuously injected with  $\eta = 0.12$  and $\gamma_e=1.9$ set equal for all pulsars. 
The  AMS-02 data for the $e^+$ flux are also shown \cite{PhysRevLett.122.041102}.}  
\label{fig:atnf_positron}
\end{figure}
The  $e^+$ flux at Earth computed for each pulsar of the ATNF catalog, older than $T>50$~kyr, is shown in Fig.~\ref{fig:atnf_positron}. 
For the diffusion around the pulsar, we have set $r_b =  120$~pc   and $D_0=7.8\times 10^{25}$~cm$^2$/s, which is the best fit value found in Ref. \cite{DiMauro:2019hwn} analyzing the TeV ICS emission around a sample of PWNe. 
We consider these numbers as a representative mean values for the inhibited diffusion around Galactic pulsars.
Out of the low diffusion bubble, $e^\pm$ are propagated in the Galaxy assuming the $D_{\rm K15}$ \cite{Kappl:2015bqa} Galactic average diffusion coefficient
(see also \cite{Manconi:2016byt}). 
For all the sources we have fixed $\eta = 0.12$  and $\gamma_e=1.9$. The value of $\eta$ is chosen in order to reach the level of AMS-02 data at few hundreds of GeV. The injection index values of $\gamma_e=1.9$ is instead suggested by the GeV-TeV analysis of known halos \cite{DiMauro:2019yvh,Shao-Qiang:2018zla}, being also in agreement with the expectations for the acceleration of $e^\pm$ pairs in PWNe \cite{Gaensler:2006ua,Amato:2013fua}. 
The total flux of $e^+$ originating from all the ATNF catalog is the sum of the fluxes from each source, and is displayed by a solid black line. The AMS-02 data  \cite{PhysRevLett.122.041102} are shown as well for comparison. 
We can note that few sources contribute around 10\% of the total measured flux at different energies. 
In particular, the sources that have a flux  $E^3 \Phi > 3\times10^{-4}$ GeV$^2$ (cm$^2$ s sr)$^{-1}$ are Geminga, B1001-47, B1055-52, B1742-30, and J1836+5925. They are all very powerful, with $\dot{E}\sim 10^{34}$~erg/s, nearby $d<0.4$~kpc and with $T$ of few hundreds of kyr. 
However, this specific list is not very informative, since it can change according to 
some parameters of our analysis (see, $e.g.$, Fig.~\ref{fig:atnf_positron_varied}).
The cumulative flux is at the level of the AMS-02 data. In particular, it can fully explain the data above 50 GeV. The conversion efficiency in to $e^\pm$, $\eta$, acts as an overall normalization factor. 
The total flux is decreasing above 1 TeV.  As discussed below, this behavior  is the result of energy losses, pulsar distance, $D_0$ and $r_b$ (being $E_c= 10^3 $ TeV).

The small features which are found in the AMS-02 data at different $E$ might be due to particular combinations of parameters for each PWNe, which are here instead considered to have all the same injection parameters $\gamma_e$ and $\eta$. 
This argument applies, in particular, for the last AMS-02 data points, for which a small variation of $\eta, \gamma_e$ for the few dominant sources can easily solve the apparent tension. 
Our main focus  is to explore the consequences of the recent results for the ICS halos for the $e^+$ flux within the two-zone diffusion model, and by varying the physical parameters connected to the inhibited diffusion zone. No attempt is made to fit the AMS-02 data points. An extensive fit of all the $e^\pm$ fluxes under this model might thus require more freedom in the specific source parameters, and is left to future work.

\begin{figure*}[]
\centering \includegraphics[width=0.49\textwidth]{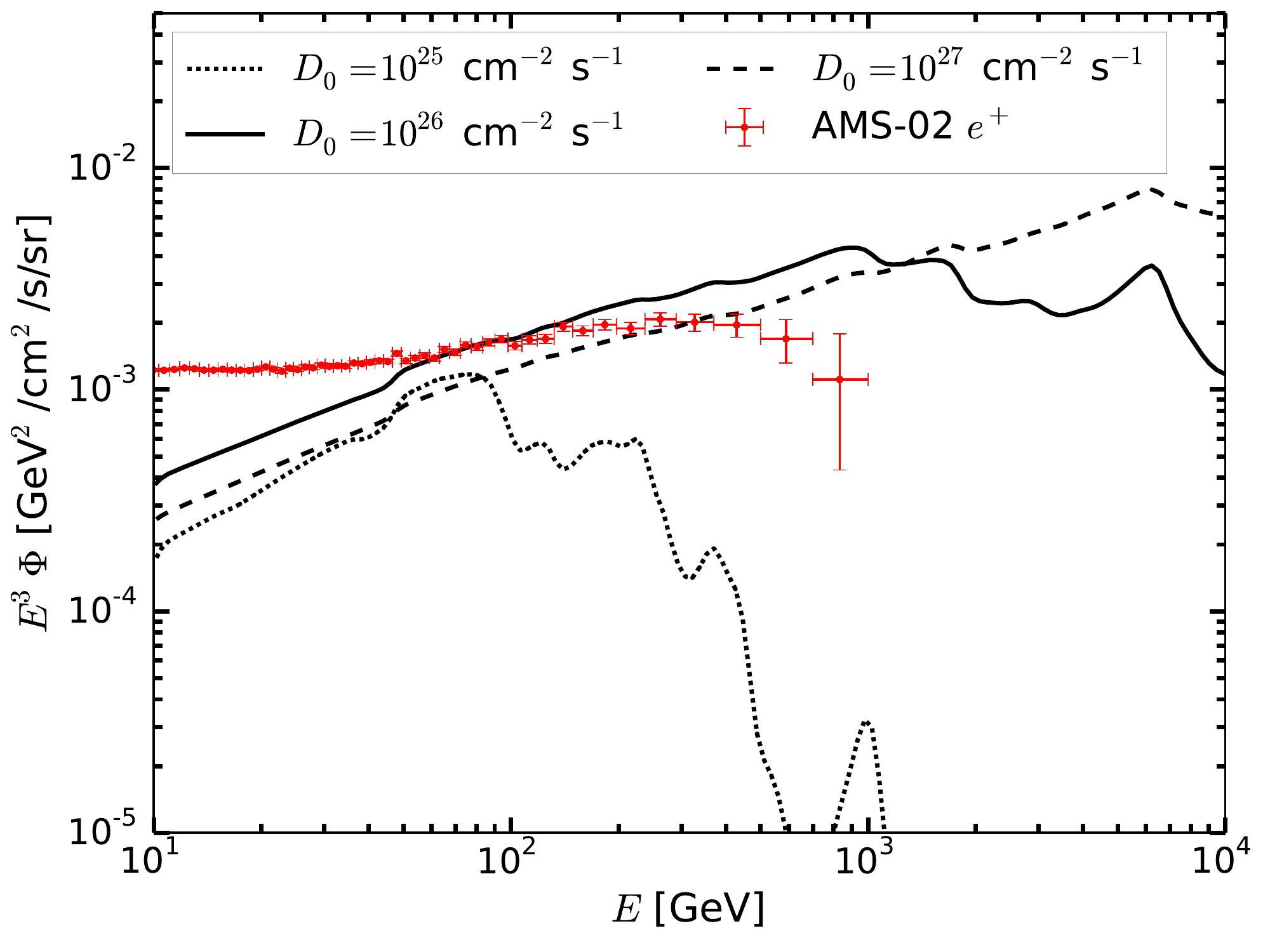}
\includegraphics[width=0.49\textwidth]{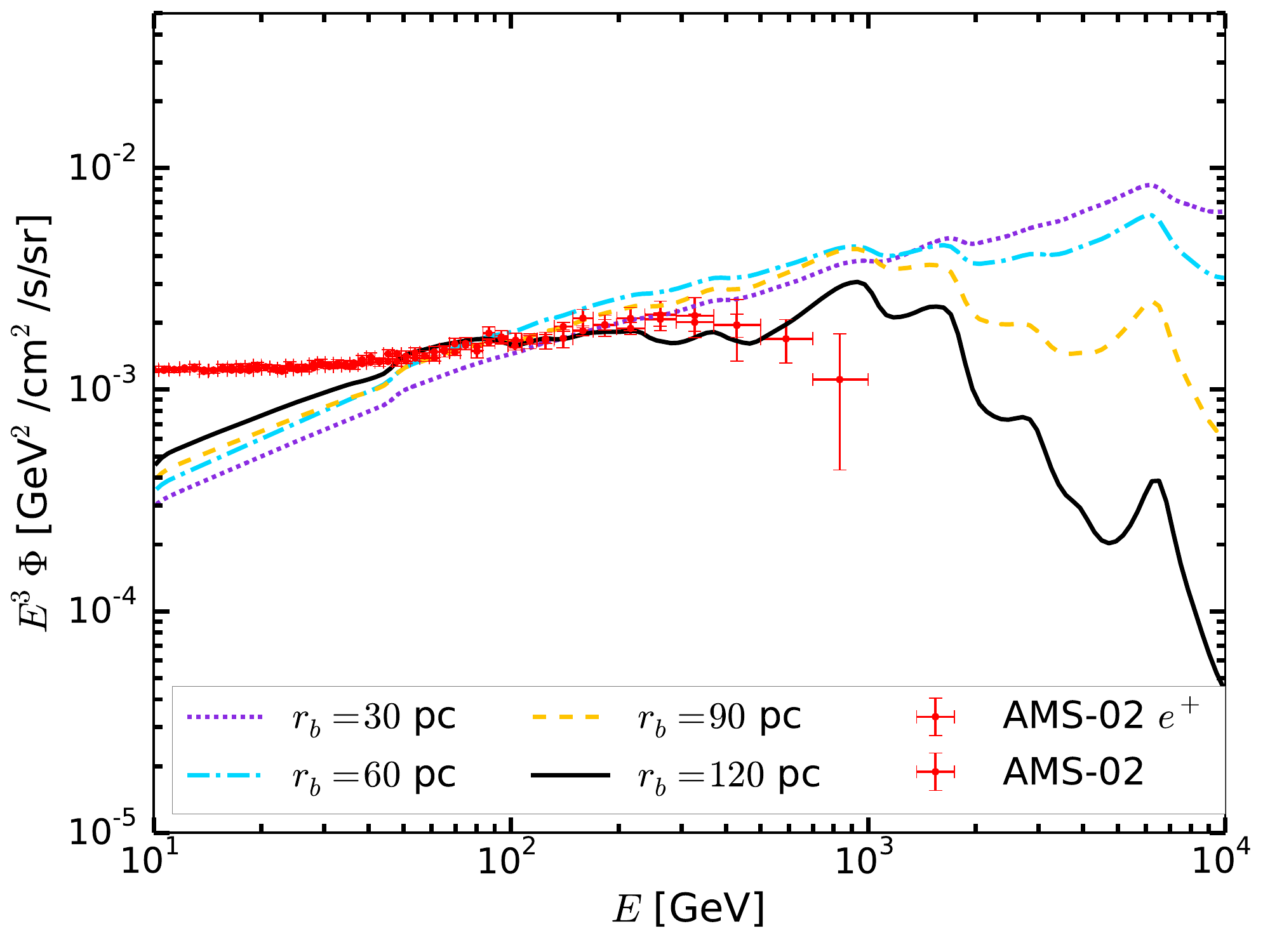}
\caption{Effect of varying $D_0$  and $r_b$  on the cumulative contribution to $e^+$ flux originating from all the ATNF pulsars. Left panel: variation of $D_0=10^{25}, 10^{26}, 10^{27}$  cm$^2$/s with fixed $r_b=90$~pc. 
Right panel: variation of $r_b = 30, 60, 90, 120$ pc with fixed $D_0=7.8\times 10^{25}$ cm$^2$/s. 
Data from AMS-02 \cite{PhysRevLett.122.041102} are reported for comparison. }  
\label{fig:rings_rb}
\end{figure*}
%
The values of $r_b$ and  $D_0$ we used in Fig.~\ref{fig:atnf_positron} have been suggested by the first observations of ICS halos, which can be considered as representative for the mean properties of all Galactic PWNe.
We have also studied the effect of the variation of $D_0$ and $r_b$ in the inhibited diffusion zone around pulsars. The results are shown in Fig.~\ref{fig:rings_rb}, for $\eta=0.12$ and $\gamma_e=1.9$. 
In the left panel, we fix $r_b=90$~pc and vary $D_0=10^{25}, 10^{26}, 10^{27}$  cm$^2$/s. An inhibited diffusion around the pulsars decreases the number of $e^+$ arriving at Earth, in particular the high energy ones. 
They are confined for a longer time before being released, therefore loose more energy. 
The mean value found in the population study in our Ref. \cite{DiMauro:2019hwn} is $D_0=7.8\times 10^{25}$ cm$^2$/s, which produces the drop of the flux above few TeV shown in Fig.~\ref{fig:atnf_positron}. 
The variation of $r_b$ = 30, 60, 90 and 120 pc  is studied in the right panel of Fig.~\ref{fig:rings_rb}. The effect  of changing the low diffusion region size around pulsars is 
correlated with the variation of $D_0$. 
By increasing $r_b$, fewer $e^+$ arrive at the Earth, in particular the high energy ones. Moving from $r_b = 60$ pc to $r_b = 120$ pc the flux decreases by about one order of magnitude at $E=10$~TeV. Below $E\simeq1$ TeV the effect of changing $r_b$ is very mild, at fixed $D_0=7.8\times 10^{25}$~cm$^2$/s. 

\begin{figure}[t]
\centering
\includegraphics[width=0.5\textwidth]{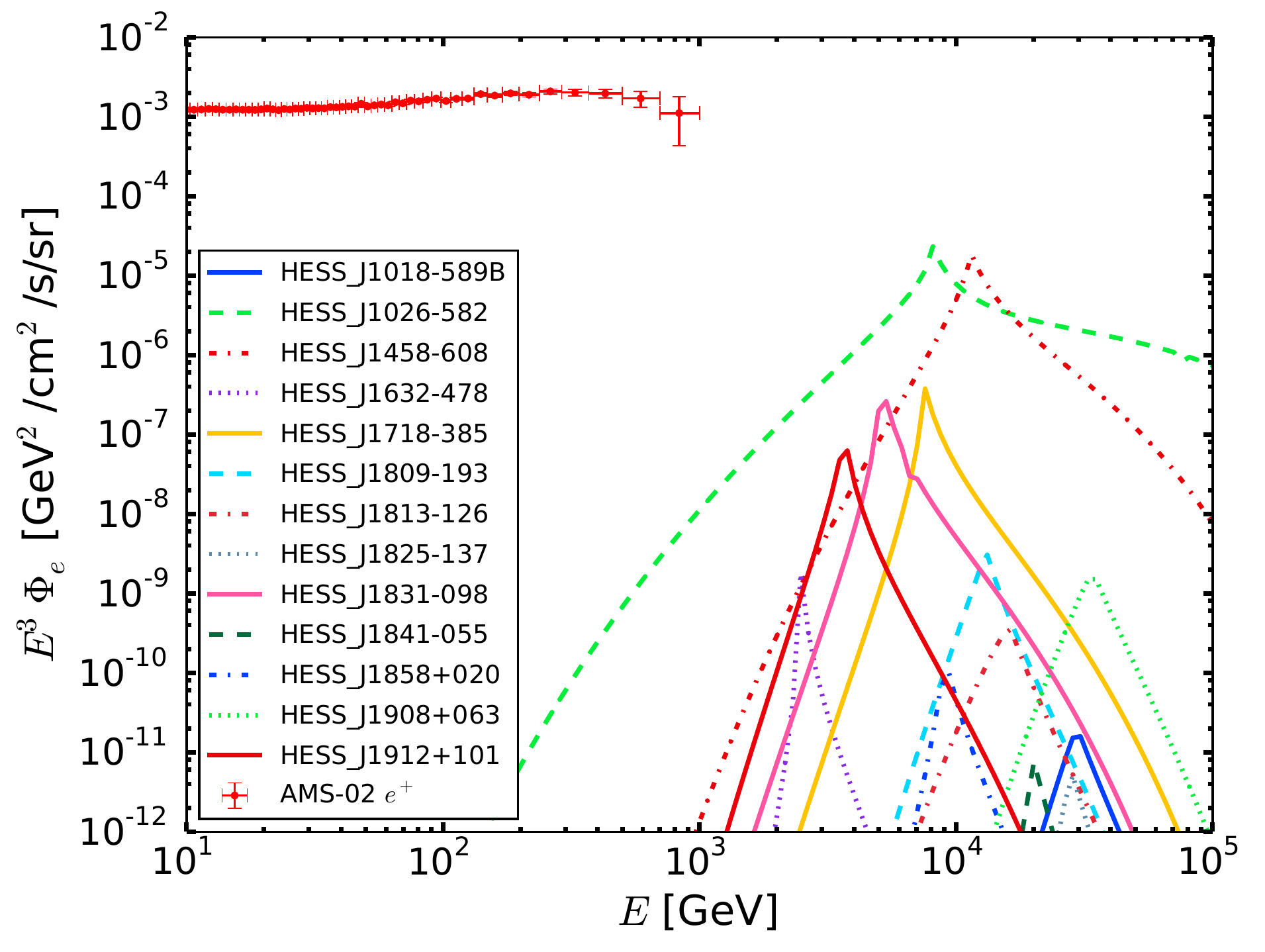}
\caption{ Predictions for the $e^+$ flux at the Earth for the PWNe in the sample of Ref. \cite{DiMauro:2019hwn}. 
The $e^+$ flux is computed by taking the values of $D_0, r_b$, $\eta$, $\gamma_e$ for each source as found in our $\gamma$-ray analysis of their ICS halos. 
AMS-02 data are also shown \cite{PhysRevLett.122.041102}.}  
\label{fig:sample_positron}
\end{figure}
We also estimate the individual contribution to the $e^+$ flux at the Earth from the specific sources studied in  \cite{DiMauro:2019hwn}, Tab.~III and IV. These sources have been observed as extended ones, and are promising candidates to possess an ICS halo. 
Here we concentrate on the sample of sources with ages larger than $20$~kyr, and compute the $e^+$ flux within the two-zone diffusion model setting 
 $D_0, r_b$, $\eta$ and $\gamma_e$ to their best fit in that analysis (see Tab.~IV in Ref. \cite{DiMauro:2019hwn}) for each source. The typical efficiency value that we have found in Ref. \cite{DiMauro:2019hwn} is of the order of $\eta \sim 0.1$ and it is compatible with the values we will use in the rest of the paper.
Here, we further select those sources with a maximum flux exceeding $10^{-15}$ GeV$^2$/cm$^2$/s/sr at least for one value of the explored energy range. 
The resulting $e^+$ fluxes are reported in Fig.~\ref{fig:sample_positron}. Within  $d<2$~kpc, the sources HESS-J1026-582 (green dashed) and HESS-1458-608 (red dotted) give the larger contribution to the $e^+$ flux in the TeV energy range. This is understood given that they are associated to the only two pulsars with $d<2$~kpc in the analyzed sample.
Nevertheless, their $e^+$ flux stands more than one order of magnitude below Geminga and Monogem at $E<5$~TeV, see \cite{DiMauro:2019yvh}.
Clearly, the sum of the $e^+$ flux produced by this the specific sample of sources  cannot explain the AMS-02 data. 
This result is not unexpected, as these sources are a small subset of the PWNe in our Galaxy.

\begin{figure}[t]
\centering
\includegraphics[width=0.5\textwidth]{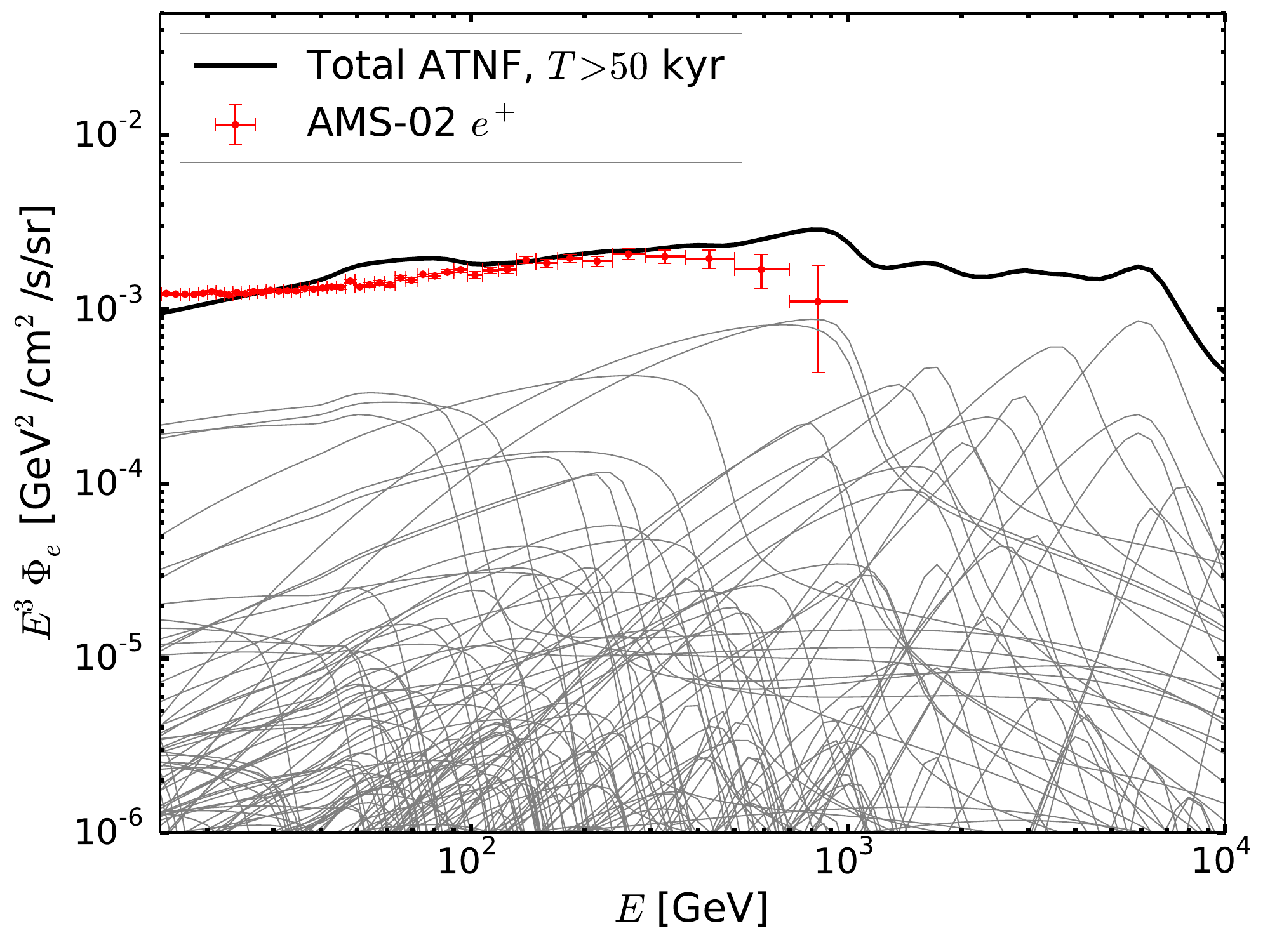}
\caption{ Same as Fig.~\ref{fig:atnf_positron}, but varying some properties of the $e^+$ emission from ATNF pulsars. 
For each source, the parameter $\gamma_e$ is set from an uniform distribution in [1.4,2.2]  and $\eta$ in [0.02, 0.30]. 
The propagation model is a two-zone diffusion model with $r_b=90$~pc, $D_0=7.8\times 10^{25}$~cm$^2$/s around each pulsar, and K15 elsewhere.}  
\label{fig:atnf_positron_varied}
\end{figure}
Finally, in order to understand the effect of   $\gamma_e$ and $\eta$, we have computed the $e^+$ flux from each ATNF source picking these parameters 
 from a uniform distribution in the range  $\gamma_e = [1.4,2.2]$  and $\eta=[0.02, 0.30]$. 
 This case is meant to mimic the variation of the injection parameters for each source.
The propagation model is a two-zone diffusion model with $r_b=90$~pc, $D_0=7.8\times 10^{25}$~cm$^2$/s around each pulsar, and K15 elsewhere. 
The results are shown in Fig.~\ref{fig:atnf_positron_varied}. 
The total flux is at the level of the data.  Given the combination of  $r_b=90$ pc  and $D_0=7.8\times 10^{25}$~cm$^2$/s, we can expect that the high energy trend in the $e^+$ flux is almost flat till  at least 10 TeV.

 \subsection{Results for simulated pulsars}
 \label{sec:res_sim}
In order to compensate the ATNF catalog incompleteness, we have performed simulations of the Galactic population of pulsars as described in Sec.~\ref{sec:simulations}. 
\begin{figure*}[t]
\includegraphics[width=0.49\textwidth]{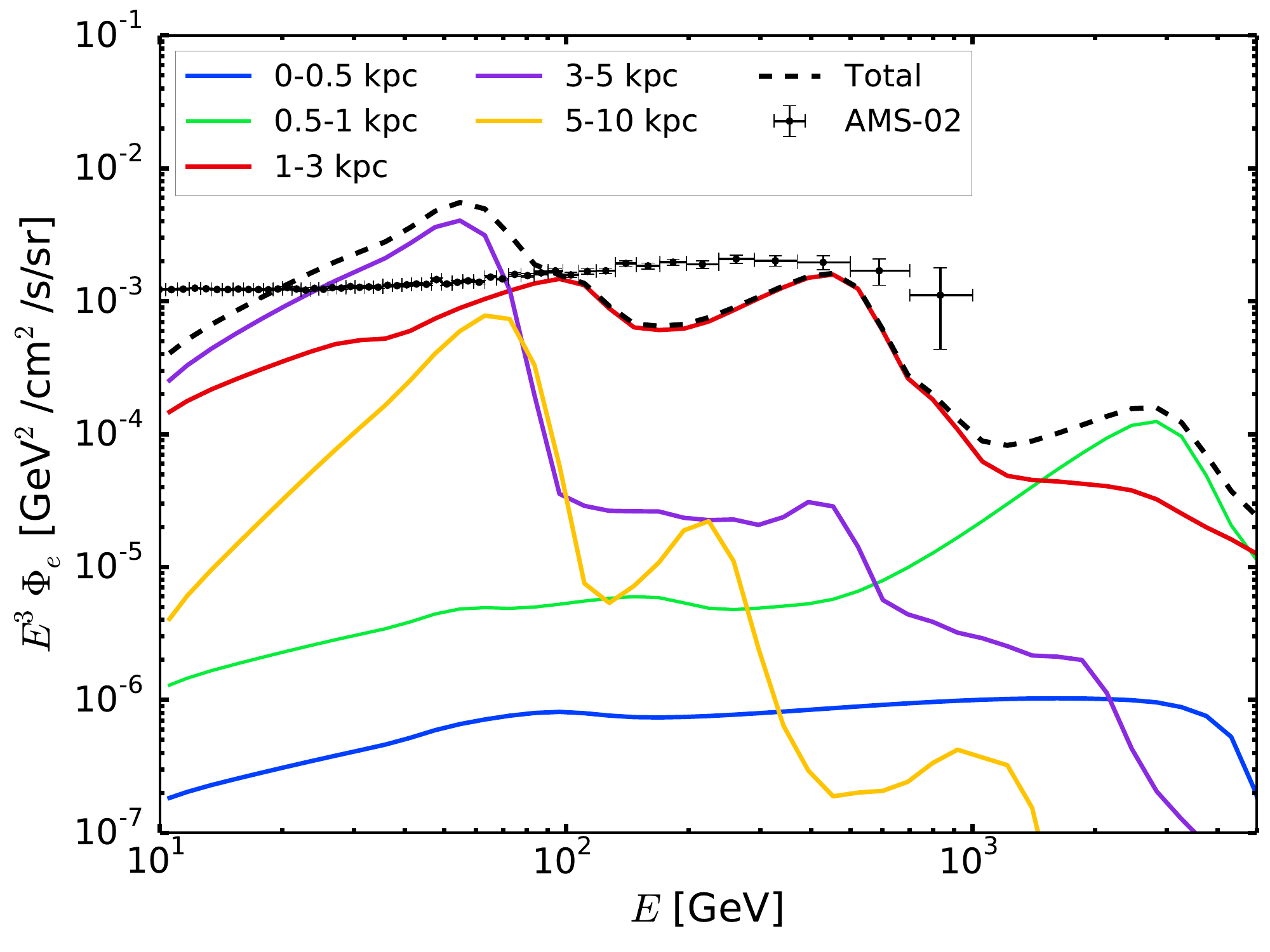}
\includegraphics[width=0.49\textwidth]{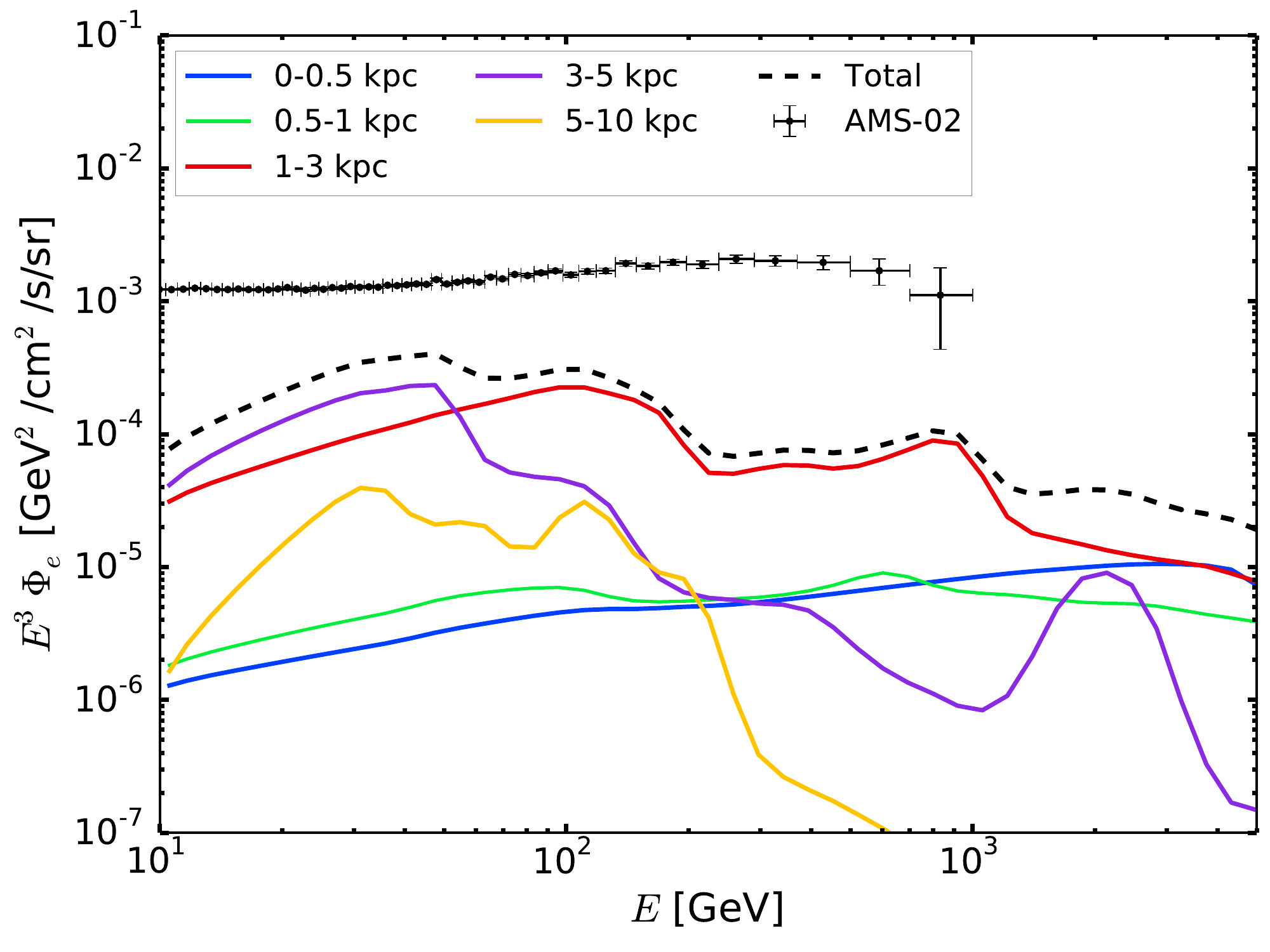}
\caption{The contribution to the $e^+$ flux at the Earth from simulated PWNe for different distance rings and for two representative simulations. The distance is intended with respect to Earth. 
 Left panel: simulated pulsars share a common  $\gamma_e=1.9$. Right panel: another realization of the Galactic pulsar distribution, where also $\gamma_e = [1.4,2.2]$.   AMS-02  data from \cite{PhysRevLett.122.041102}. }
\label{fig:rings}
\end{figure*}
For a specific Galactic pulsar realization, we have computed the $e^+$ flux at Earth fixing $r_b=90$~pc and $D_0=7.8\times 10^{25}$~cm$^2$/s for the inhibited diffusion zone, the K15 parameters for the rest of the Galaxy, $\eta=0.06$, and selecting ages larger than $20$~kyr. 
We show the result for this illustrative simulation in Fig.~\ref{fig:rings}, where the flux of $e^+$ from each source has been summed in separate  rings of distance from the Earth.
In the left panel, fluxes are computed for  $\gamma_{\rm e}=1.9$. In the right panel, for another Galactic realization, $\gamma_e$ is picked from a uniform distribution in the range  [1.4,2.2]. 
In both panels, the black dashed line is the sum for all the simulated sources at $d<10$~kpc, while the colored solid lines indicate the contribution for each distance ring.
A general comment to the figure is that the flux at Earth for $E>100$~GeV is dominated by  sources at $d<3$~kpc. 
This is understood given the typical propagation length of high-energetic $e^+$, affected by severe energy losses.
Despite the small energy losses, the flux from sources within 1 kpc from the Earth is low due to the paucity of sources. 
Instead, the ring $1-3$~kpc contributes significantly to the total flux since the presence of a spiral arm enhances the number of sources. 
The contribution of sources between $3$~kpc and $5$~kpc changes the total flux at the percent level at $E>100$~GeV, while it gives the dominant contribution for $E<100$~GeV.
The $e^+$ flux from the sources in the distance range of $5-10$~kpc is negligible for $E>100$~GeV, while at lower energies is can range 10\% of the total at a specific energy and only in the simulation shown in the left panel. 
Indeed, the relative contribution of the distance rings to the total flux  depends on the particular simulation realization. This is mainly due to the fact that  the flux is often dominated by few powerful sources, in particular for $E>500$~GeV, as it is visible from the peaks in the flux in Fig.~\ref{fig:rings}.
Nevertheless, we checked by means of twenty simulations that the  $1-3$~kpc ring gives always the dominant contribution to the total $e^+$ flux at high energies. 
We also verified that a change of $r_b$ and $D_0$ leads to similar conclusions for the relative contributions from different distance rings to the total flux, while it can change significantly the total flux at Earth, as previously shown fro the ATNF cataloged pulsars. 
The AMS-02 data for the $e^+$ flux are shown in Fig.~\ref{fig:rings} for comparison (no fit has been performed).  
The total flux of $e^+$ for the illustrative left panel simulation is at the level of AMS-02 data up to a factor of five in all the energy range, with an overshooting of the data at about 50 GeV. 
The simulation reported in the right panel instead corresponds to a total $e^+$ flux smaller than the AMS-02 data. Note again that the efficiency has always been set to $\eta=0.06$. 

\begin{figure*}[t]
\centering \includegraphics[width=0.49\textwidth]{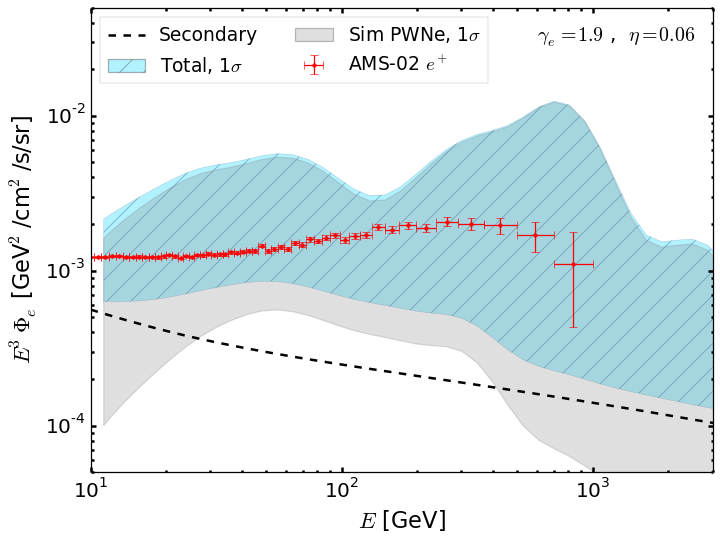}
\centering \includegraphics[width=0.49\textwidth]{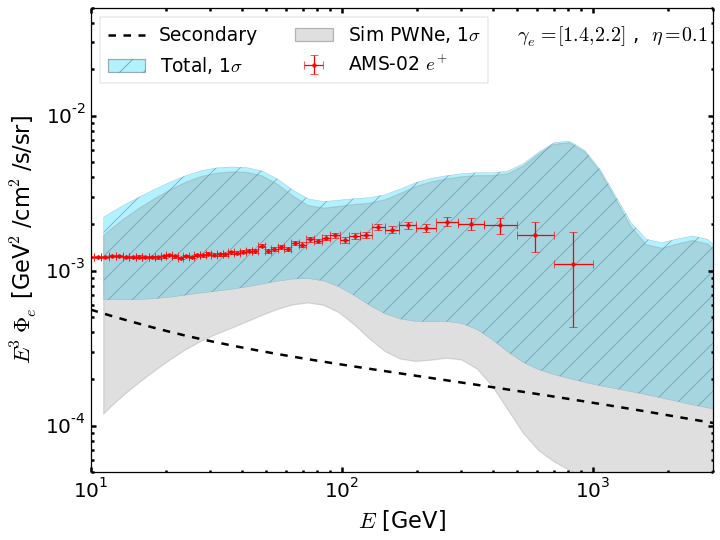}
\caption{Flux of  cosmic-ray $e^+$ flux at Earth obtained from simulated PWN populations (grey band), from the  secondary emission (black dashed line), and by the sum of the two contributions (cyan hatched band). The bands are computed by  considering the $1\sigma$ deviation from the mean of ten  simulations, using a two-zone diffusion model with $D_0=7.8\times10^{25}$cm$^2$/s, $r_b=90$~pc. 
Left  panel: $\gamma_e=1.9$ and $\eta=0.06$; right panel: uniform distribution of $\gamma_e$ in the range [1.4, 2.2] and $\eta=0.1$. 
AMS-02 data from \cite{PhysRevLett.122.041102}. }  
\label{fig:20sim_mean}
\end{figure*}
A more reliable validation of our model can be drawn only after several simulations. The different source realizations  affect in particular the flux at the highest energies,
which are dominated by local sources. 
We have performed ten simulations of the Galactic PWN population, computed the total $e^+$ flux at the Earth for each realization, and derived the mean flux and its $1\sigma$ standard deviation. 
We consider for each mock source $D_0=7.8\times10^{25}$cm$^2$/s, $r_b=90$~pc, and K15 propagation for $r>r_b$, while the efficiency $\eta$ is tuned in order to give a mean total $e^+$ flux at the level of AMS-02 data, for a pulsar rate of 1 per century per Galaxy. 
We have added, to the contribution from simulated PWNe, the secondary $e^+$, which are computed within the K15 model as done in \cite{DiMauro:2019yvh,Manconi:2018azw,DiMauro:2017jpu}. 
In Fig.~\ref{fig:20sim_mean} we show the flux of  cosmic-ray $e^+$ flux at Earth obtained from simulated PWN populations (grey band), from the  secondary emission (black dashed line), and by the sum of the two contributions (cyan hatched band). 
The grey band indicates the contribution from PWNe for $1\sigma$ from the mean value, computed in logarithmic scale, of the ten simulations, while the cyan hatched band is the sum of the PWNe contribution and the secondary emission. 
The left panel reports the case of fixed  $\gamma_e=1.9$ and $\eta=0.06$, the right panel the case for $\gamma_e$ picked from a uniform distribution  in the range [1.4, 2.2]. 
The differences within each PWNe realization cause the $1\sigma$ band to be at least one order of magnitude in all the 
energy range. The different intensity and position of the flux peaks is due to the different realization of few, powerful and nearby sources, and their different $\gamma_e$ values and $\eta$. 
A hint for a decreasing $e^+$ flux at TeV energies is found, even if this is strongly connected to the chosen value of $r_b$, see Fig.~\ref{fig:rings_rb}.
It can be realized in a two-zone diffusion model, provided the radius of the low diffusion bubble in sizable ($\gsim$ 100 pc) and the diffusion coefficient inside the bubble is small ($D_0 \lsim 6-7 \times10^{25}$cm$^2$/s). Given the $\gamma$-ray observation of several PWNe at energies well above the TeV \cite{Abeysekara:2017science,Abeysekara:2019gov}, it is hard to hypothesize 
a cut-off in the $e^\pm$ source spectrum (see Eq. \ref{eq:Q_E_cont}). A detailed analysis of these properties if left to a future work.
We  have also studied the effect of an age lower cut to 50 kyr, as done for the ATNF pulsars. The differences are not significant, being smaller than $10^{-2}-10^{-3}$ on the whole energy spectrum. The 
sources with 20 kyr$<T<$ 50 kyr are scarce, and even less are  the close ($d<$ 5 kpc) and powerful ones. 

The simulated sources with a common spectral index $\gamma_e=1.9$ and $\eta=0.06$ lead to similar results of the simulations with variable $\gamma_e$, within $1\sigma$  of uncertainty, provided that $\eta$ is set to 0.1. 
Remarkably, the total contribution from the PWNe and the secondary $e^+$ is at the level of AMS-02 data for an efficiency of conversion of the pulsar spin down energy in $e^-$ and $e^+$ pairs of $\eta\sim0.1$. 
We note however that this number is fixed equal for all the sources, which is a reasonable assumption in a population study like the one we have conducted, but can realistically be different from source to source. 
Compared to the values for $\eta$ required for the ATNF PWNe case (see Sec.~\ref{sec:res_cat}) to reach the level of AMS-02 data, we find that the $\eta$ required for the simulated sample of Galactic PWNe is systematically lower, and more similar to the values which are found for Geminga and Monogem \cite{DiMauro:2019yvh} using the same $\gamma_e=1.9$. This is expected, as a higher value for the efficiency might be compensating some level of incompleteness of the ATNF catalog.

Finally, we note that for the models investigated in this paper the dipole anisotropy in the $e^+$ or $e^+ +e^-$ fluxes is expected to be well below the current upper limits from AMS-02 \cite{PhysRevLett.122.041102} and \textit{Fermi}-LAT \cite{Abdollahi:2017kyf}. 
In fact, as extensively discussed in Refs.~\cite{Manconi:2016byt, Manconi:2018azw}, when the global contribution of all Galactic pulsars is taken into account, and there is not a single, dominating source to the $e^+$ flux, even the maximum anisotropies for the few most powerful sources are predicted to be at least one order of magnitude below current anisotropy upper limits.

\section{Conclusions}
\label{sec: conclusions}
The present paper contains a novel and extensive analysis of the contribution of the Galactic PWNe to the flux of cosmic  $e^+$. 
The main motivation for this work comes from the recent idea that inefficient diffusion zones are present around each Galactic PWNe, as emerged firstly from the  Geminga and Monogem $\gamma$-ray data measured by HAWC above TeV energy  \cite{Abeysekara:2017science}, then confirmed  around Geminga at  GeV energies in the {\it Fermi}-LAT data \cite{DiMauro:2019yvh}, and as suggested for a list of candidates in   
$\gamma$-ray halos \cite{DiMauro:2019hwn}. The key idea is that the  $\gamma$-ray halos are due to ICS of higher energy $e^\pm$ off the ISRF populations. 

Here we implement a two-zone diffusion model for the propagation of $e^+$ from each PWN, where diffusion is inhibited within a radius $r_b$ from the PWN, and takes the average ISM value elsewhere. 
We firstly apply our model to all the pulsar listed in the ATNF catalog with an age between $50$~kyr and $10^5$~kyr. 
We find that  few sources contribute around 10\% of the total measured flux at different energies and that the cumulative flux is at the level of the AMS-02 data. In particular, it can fully explain the data above 50 GeV 
with an efficiency for the conversion of the pulsar spin down energy into  $e^-$ and $e^+$ pairs of $\eta\sim0.1$
Our conclusions depend considerably on the value of the diffusion coefficient in the bubble around the pulsar, as well as from its radius, and are corroborated by the recent findings about Monogem and, particularly, Geminga PWN. 

In order to compensate the ATNF catalog incompleteness, we build up a number of simulations of the PWN Galactic population and compute the $e^+$ at Earth accordingly. 
We can therefore determine the mean flux resulting from all the sources in each single simulation, and derive the relevant $1\sigma$ band. 
The differences within each PWNe realization cause the $1\sigma$ band to be at least one order of magnitude from 10 GeV up to few TeV. 
A hint for a decreasing $e^+$ flux at TeV energies is found, even if this is strongly connected to the chosen value of the radius $r_b$ for the low diffusion zone around the sources. 
Remarkably, the total contribution from the PWNe and the secondary $e^+$ is at the level of AMS-02 data for an efficiency of conversion of the pulsar spin down energy in $e^-$ and $e^+$ pairs of $\eta\sim0.1$. 

We conclude that the global contribution from Galactic PWNe, as computed within a two-zone diffusion model, and including our constraints for the inhibited diffusion zone around each source, remains a viable interpretation for the $e^+$ flux observed by AMS-02. 
The hint of a cut-off in the predicted $e^+$ flux is only possible in a two-zone diffusion model, and with particular combinations of  $r_b$ and $D_0$.
The models discussed in this work could be tested by forthcoming new data on $e^+$ and $e^-$ fluxes, such as further statistics from the AMS-02 experiment, or by proposed missions such as the AMS-100  \cite{Schael:2019lvx} or Aladino \cite{Aladino}, as well as from further multi-wavelength analysis of the extended halos around Galactic pulsars.

\begin{acknowledgments}
This research has made use of Gammapy, a community-developed, open-source Python package for $\gamma$-ray astronomy \cite{2015ICRC...34..789D}.
The authors thank Christoph Deil and Axel Donath for the \texttt{gammapy} support.
The work of FD and SM  has been supported by the "Departments of Excellence 2018 - 2022" Grant awarded by
the Italian Ministry of Education, University and Research (MIUR) (L. 232/2016).
FD and SM acknowledge financial contribution from the agreement ASI-INAF
n.2017-14-H.0 and the Fondazione CRT for the grant 2017/58675.
\end{acknowledgments}

\bibliography{paper}

\end{document}